\documentclass[11pt,onecolumn,dvips,draftcls]{IEEEtran}

\usepackage{amsfonts,amsmath,graphicx,epsfig}
\usepackage{subfigure,graphics,amssymb,amsxtra,color}
\usepackage{algorithm}
\usepackage{algorithmic}
\usepackage[ansinew]{inputenc}

\newtheorem{cor}{Corollary}
\newcommand{\prob}{{\cal P}}
\newcommand{\pout}{{\cal P}_{\textrm{out}}}
\newcommand{\cout}{C_{\textrm{out}}}

        \newtheorem{theorem}{Theorem}%
        \newtheorem{lemma}{Lemma}%

\DeclareOldFontCommand{\rm}{\normalfont\rmfamily}{\mathrm}
\DeclareOldFontCommand{\sf}{\normalfont\sffamily}{\mathsf}
\DeclareOldFontCommand{\tt}{\normalfont\ttfamily}{\mathtt}
\DeclareOldFontCommand{\bf}{\normalfont\bfseries}{\mathbf}
\DeclareOldFontCommand{\it}{\normalfont\itshape}{\mathit}
\DeclareOldFontCommand{\sl}{\normalfont\slshape}{\@nomath\sl}
\DeclareOldFontCommand{\sc}{\normalfont\scshape}{\@nomath\sc}

\newcommand{\comment}[1]{}

\newcommand{\bml}[1]{\begin{multline}\label{#1}}
\newcommand{\eml}{\end{multline}}
\newcommand{\beq}[1]{\begin{equation}\label{#1}}
\newcommand{\eeq}{\end{equation}}

\newcommand{\beann}{\begin{eqnarray*}}
\newcommand{\eeann}{\end{eqnarray*}}
\newcommand{\bea}[1]{\begin{eqnarray}\label{#1}}
\newcommand{\eea}{\end{eqnarray}}
\newcommand{\bmp}{\begin{minipage}}
\newcommand{\emp}{\end{minipage}}

\newcommand{\eq}[1]{\eqref{#1}}

\newcommand{\secref}[1]{Section~\ref{#1}}

\newcommand{\fig}[1]{{\itshape Fig.~\ref{#1}}}

\def\SET0N {I\hspace{-0.8ex}N_0}

{%
}%

{%
 \newtheorem{remark}{Remark}%
}%
\newsavebox{\Citname}

\newcommand{\ignore}[1]{}

\newcommand{\expect}[1]{\ensuremath{\operatorname{E}\left[#1\right]}}

\newcommand{\eg}{e.g.~}

\begin{document}

\title{Wireless Information-Theoretic Security\\
--- Part I: Theoretical Aspects
\footnote{
Matthieu Bloch and Steven W. McLaughlin are with GT-CNRS UMI 2958, Metz, France, and also with the School of ECE, Georgia Institute of Technology, Altlanta, GA.\\
\indent Jo\~ao Barros is with the Departament of Computer Science \& LIACC/UP,
Universidade do Porto,  Portugal.\\
\indent Miguel R. D. Rodrigues is with the Computer Laboratory, University of Cambridge, United Kingdom.\\
\indent Parts of this work have been presented at the IEEE International Symposium on Information Theory 2006~\cite{Barros-Rodrigues:06}, at the 44th Allerton conference on Communication Control and Computing~\cite{Bloch2006b}, and at the IEEE Information Theory Wokshop 2006 in Chengdu~\cite{Bloch2006c}.}
}
\author{\authorblockN{Matthieu Bloch
, Jo\~ao Barros
, Miguel R. D. Rodrigues
, and Steven W. McLaughlin
}
} \maketitle

\begin{abstract}
In this two-part paper, we consider the transmission of confidential  
data over wireless wiretap channels.
The first part presents an information-theoretic problem formulation  
in which two legitimate partners
communicate over a quasi-static fading channel and an eavesdropper  
observes their transmissions through
another independent quasi-static fading channel. We define the  
secrecy capacity in terms of outage
probability and provide a complete characterization of the maximum  
transmission rate at which the
eavesdropper is unable to decode any information. In sharp contrast  
with known results for Gaussian
wiretap channels (without feedback), our contribution shows that in  
the presence of fading information-
theoretic security is achievable even when the eavesdropper has a  
better average signal-to-noise ratio
(SNR) than the legitimate receiver — fading thus turns out to be a  
friend and not a foe. The issue of
imperfect channel state information is also addressed. Practical  
schemes for wireless information-theoretic
security are presented in Part II, which in some cases comes close to  
the secrecy capacity limits given in this paper.
\end{abstract}
\newpage

\section{Introduction}

The issues of privacy and security in wireless communication networks  
have taken on an increasingly
important role as these networks continue to flourish worldwide.  
Traditionally, security is viewed as an
independent feature addressed above the physical layer and all widely  
used cryptographic protocols (e.g. RSA, AES etc) are designed and  
implemented assuming the physical layer has already been established  
and is error free.

\begin{figure}[b!]
  \centering
  \includegraphics[width=7.5cm]{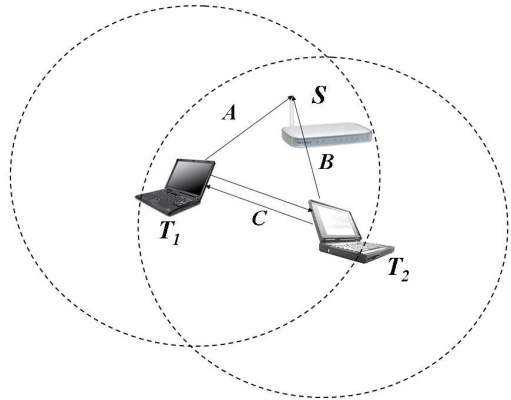}
  \caption{Example of a wireless network with potential eavesdropping. Terminals $T_1$ and $T_2$
  communicate with a base station $S$ over a wireless medium (channels $A$ and $B$).
  By listening to the transmissions of terminal $T_1$ (through channel $C$),
  terminal $T_2$ may acquire confidential information.
  If $T_1$ wants to exchange a secret key or guarantee the confidentiality
  of its transmitted data, it can
  exploit the {\it physical} properties of the wireless channel
  to secure the information by {\it coding} against Terminal $T_2$.}
  \label{fig:wirelessnet}
\end{figure}

In contrast with this paradigm, there exist both theoretical and
practical contributions
that support the potential of physical layer security ideas to
significantly strengthen the security of digital communication
systems. The basic principle of {\it information-theoretic
security} --- widely accepted as the strictest notion of security
--- calls for the combination of cryptographic schemes with
channel coding techniques that exploit the randomness of the
communication channels to guarantee that the sent messages cannot
be decoded by a third party  maliciously eavesdropping on the
wireless medium (see \fig{fig:wirelessnet}).

The theoretical basis for this information-theoretic approach, which
builds on Shannon's notion of {\it perfect
secrecy}~\cite{Shannon:BSTJ49}, was laid by
Wyner~\cite{Wyner:BSTJ75} and later by Csisz\'ar and
K\"orner~\cite{Csiszar&Korner:IT78}, who proved in seminal papers
that there exist channel codes guaranteeing both robustness to
transmission errors and a prescribed degree of data confidentiality.

\begin{figure}[t!]
  \centering
  \includegraphics[width=13cm]{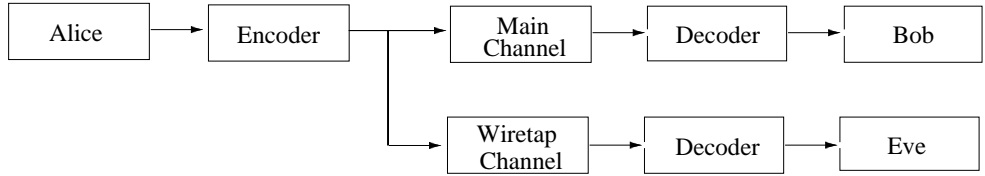}
  \caption{In the wiretap channel problem, the goal of the legitimate users, Alice and Bob, is to communicate reliably over the noisy main channel, while ensuring that
  an eavesdropper, say Eve, is unable to obtain any information from the outputs of the wiretap channel.}
  \label{fig:wiretap}
\end{figure}

A general setup for the so called wiretap channel is shown in~\fig{fig:wiretap}. In the original version, proposed by Wyner in~\cite{Wyner:BSTJ75},
two legitimate users communicate over a main channel and an
eavesdropper has access to degraded versions of the channel
outputs that reach the legitimate
receiver. In \cite{CheongH:78} it was shown that if both the main
channel and the wiretap channel are additive white Gaussian noise
(AWGN) channels, and the latter has less capacity than the former,
the {\it secrecy capacity} (i.e. the maximum transmission rate at
which the eavesdropper is unable to decode any information) is equal
to the difference between the two channel capacities. Consequently,
confidential communication is not possible unless the Gaussian main
channel has a better signal-to-noise ratio (SNR) than the Gaussian
wiretap channel.

In the seventies and eighties, the impact of these works was limited,
partly because practical wiretap codes were not available, but mostly
because due the fact that a strictly positive secrecy capacity in the
classical wiretap channel setup requires the legitimate sender
and receiver to have some advantage (a better SNR) over
the attacker. Moreover, almost at the same time, Diffie and
Hellman~\cite{Diffie:76a} published the basic principles of
public-key cryptography, which was to be adopted by nearly all
contemporary security schemes.

More recently, information-theoretic security witnessed a renaissance
arguably due to the work of Maurer~\cite{Maurer:93}, who proved that
even when the legitimate users (say Alice and Bob) have a worse channel
than the eavesdropper (say Eve), it is possible for them to generate
a secret key through public communication over an insecure yet
authenticated channel. In~\cite{MaurerW:00} Maurer and Wolf showed that a
stronger (and technically more convincing) secrecy condition for discrete
memoryless channels yields the
same secrecy rates as the weaker condition in ~\cite{Wyner:BSTJ75}
and ~\cite{Csiszar&Korner:IT78}.
A key ingredient for secret key generation over
noisy channels is privacy amplification (see Bennett et al~\cite{BBCM95}),
which provides Alice and Bob with the means to distill perfectly secret
symbols (e.g.~a secret key) from a large set of only partially
secret data. This general approach is used and
modified in Part II of this paper to develop efficient protocols for the
Gaussian and quasi-static fading wiretap channel.

In~\cite{hero:01}, Hero introduced space-time signal processing
techniques for secure communication over wireless links.
More recently, Parada and Blahut~\cite{Parada2005} considered the
secrecy capacity of various degraded fading channels.
In a shorter prelude to some of the results in this paper
~\cite{Barros-Rodrigues:06},
Barros and Rodrigues provided
the first characterization of the outage secrecy capacity of slow
fading channels and showed that in the presence of fading
information-theoretic security is achievable even when the eavesdropper
has a better average signal-to-noise ratio (SNR) than the
legitimate receiver--- without the need
for public communication over a feedback channel. The ergodic secrecy
capacity of fading channels was soon derived  by Liang and Poor
~\cite{Liang2006}, and, independently, by Li et al.~\cite{LiYT:06}.
Power and rate allocation schemes for secret communication over fading channels
were presented by Gopala et al. in ~\cite{gopala-2006}. Secure
broadcasting over wireless channels is considered in~\cite{Khisti2006}.

Practical secrecy capacity-achieving codes for erasure channels were
presented by Thangaraj et al. in ~\cite{Thangaraj:04}. LDPC codes
were also shown by Bloch et al.~\cite{BlochTM:05} to be useful tools
for reconciliation of correlated continuous random variables,
with implications in quantum key distribution. A related scheme
was presented by Ye and Reznik in~\cite{Ye2006}. Experimental results
supporting the possibility of information-theoretic
secret key agreement over wireless channels were reported by Imai
et al in~\cite{Imai2006}.

Secrecy systems with multiple users have also recently become an object
of intense research. Csiszár and Narayan~\cite{Csiszar2004}
presented the fundamental limits of secret key generation
in multi-terminal setups. Secret key constructions
for this problem are reported by Ye and Narayan in~\cite{Ye2006a}.
A detailed study of the multiple access channel with secrecy constraints
between users
was provided by Liang and Poor in~\cite{Liang2006a}. Liu et al
presented results for the same problem in~\cite{Liu2006a} and,
investigated in~\cite{Liu2006}  also broadcast and
interference channels with confidential messages.
The Gaussian multiple access channel with an eavesdropper
was studied in~\cite{Tekin2006}.
\subsection{Our Contributions}

Motivated by the general problem of securing transmissions over
wireless channels, we consider the impact of fading on the secrecy
capacity. Our contributions in Part I are as follows:
\renewcommand{\labelenumi}{(\alph{enumi})}
\begin{enumerate}
\item an
information-theoretic formulation of the problem of secure
communication over wireless channels;
\item a characterization of the
secrecy capacity of single-antenna quasi-static Rayleigh fading
channels in terms of outage probability;
\item a simple analysis of
the impact of user location on the achievable level of secrecy;
\item
a rigorous comparison with the Gaussian wiretap channel evidencing
the benefits of fading towards achieving a higher level of security;
\item a mathematical characterization of the impact of imperfect
CSI about the eavesdropper's channel on the secrecy capacity;
\item a comparison between information-theoretic security techniques
at the physical layer and classical cryptographic methods at higher
layers of the protocol stack.
\end{enumerate}

Among the different conclusions to be drawn from our results perhaps
the most striking one is that, in the presence of fading,
information-theoretic security is achievable even when the eavesdropper's
channel has a better average SNR than the main channel.

\subsection{Organization of the Paper}

The rest of the paper is organized as follows. First,
\secref{sec:statement} provides an information-theoretic
formulation of the problem of secure communication over fading
channels. Then, \secref{sec:outage} analyzes the secrecy capacity of a
quasi-static Rayleigh fading channel in terms of outage
probability. The implications of channel state information
are analyzed in \secref{sec:csi}. Finally, \secref{sec:comparison}
compares classical cryptographic methods with information-theoretic
security for wireless channels, and \secref{sec:conclusions}
concludes the paper.

\section{Secure Communication over Quasi-Static Rayleigh Fading Channels}
\label{sec:statement}

\subsection{Wireless System Setup}

\begin{figure}[t!]
  \centering
  \includegraphics[width=12cm]{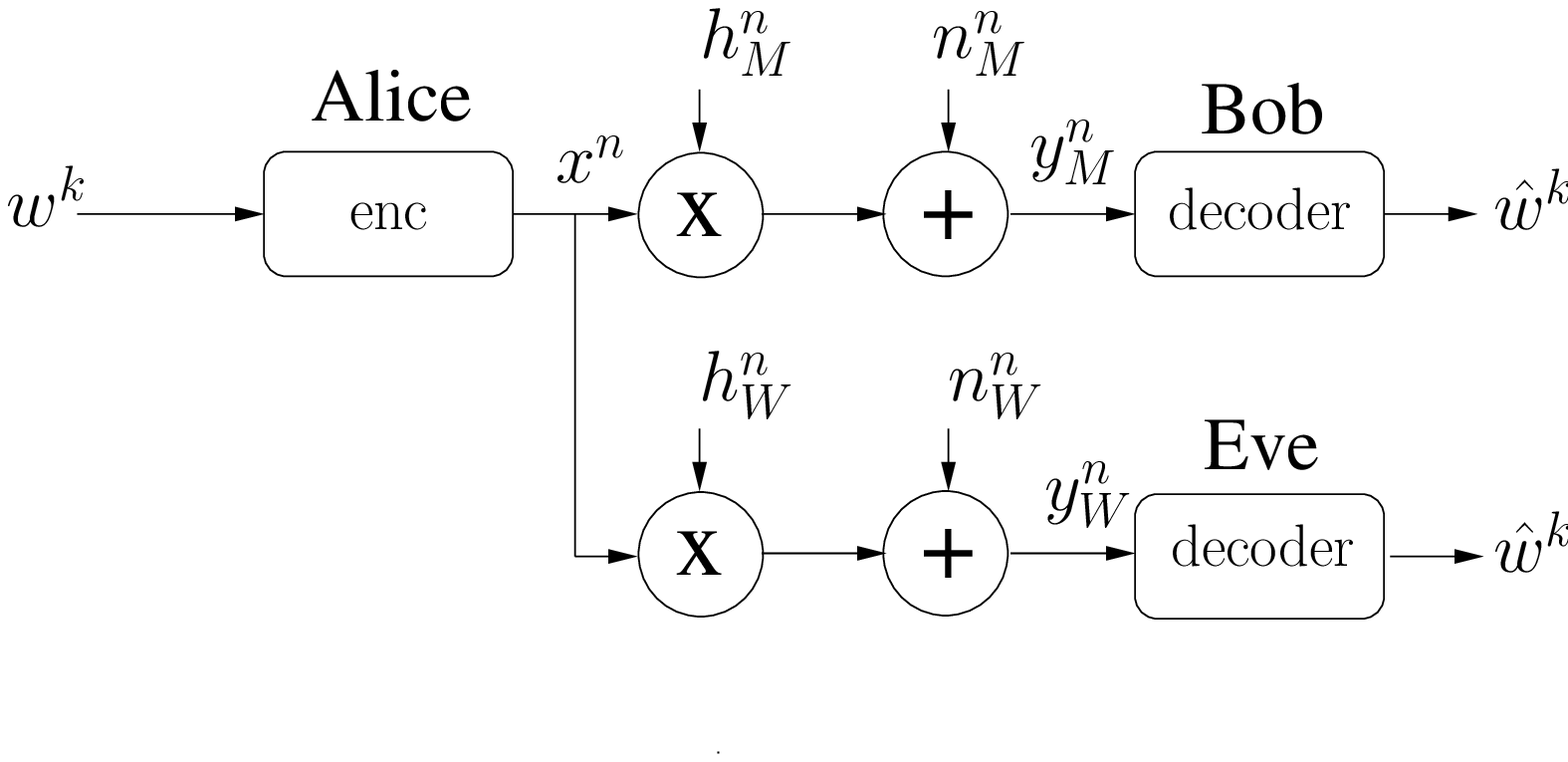}
  \caption{Wireless system setup. }
  \label{fig:wireless_system_setup}
\end{figure}

Consider the wireless system setup depicted in \fig{fig:wireless_system_setup}. A legitimate user, say Alice, wants
to send messages to another user, say Bob. Alice encodes the message
block $w^k$ into the codeword $x^n$ for transmission over the
channel (the {\it main} channel). Bob observes the output of a
discrete-time Rayleigh fading channel given by
\[y_M(i)=h_M(i)x(i)+n_M(i),\]
where $h_M(i)$ is a circularly symmetric complex Gaussian random
variable with zero-mean and unit-variance representing the main
channel fading coefficient and $n_M(i)$ is a zero-mean circularly
symmetric complex Gaussian noise random variable.

A third party (Eve) is also capable of eavesdropping Alice's
transmissions. In particular, Eve observes the output of an
independent discrete-time Rayleigh fading channel (the {\it wiretap}
channel) given by
\[y_W(i)=h_W(i)x(i)+n_W(i),\]
where $h_W(i)$ denotes a circularly symmetric complex Gaussian
random variable with zero-mean and unit-variance representing the
wiretap channel fading coefficient and $n_W(i)$ denotes a zero-mean
circularly symmetric complex Gaussian noise random variable.

It is assumed that the channels' input, the channels' fading
coefficients and the channels' noises are all independent. It is also
assumed that both the main and the wiretap channels are quasi-static
fading channels, that is, the fading coefficients, albeit random,
are constant during the transmission of an entire codeword
($h_M(i)=h_M, \forall i=1,\ldots,n$ and $h_W(i)=h_W, \forall
i=1,\ldots,n$) and, moreover, independent from codeword to codeword.

We take the average transmit power to be $P$, that is
\[ \frac{1}{n} \sum_{i=1}^n \expect{|X(i)|^2} \leq P, \]
and the average noise power in the main and the wiretap channels to
be $N_M$ and $N_W$, respectively. Consequently, the instantaneous
SNR at Bob's receiver is
\[\gamma_M(i) = P|h_M(i)|^2/N_{M} = P|h_M|^2/N_{M} = \gamma_M\]
and its average value is
\[\overline \gamma_M(i)=P \expect{|h_M(i)|^2}/N_{M} = P \expect{|h_M|^2}/N_{M} = \overline \gamma_M.\]
Likewise, the instantaneous SNR at Eve's receiver is
\[\gamma_W(i) = P|h_W(i)|^2/N_{W} = P|h_W|^2/N_{W} = \gamma_W\]
and its average value is
\[\overline \gamma_W(i)=P \expect{|h_W(i)|^2}/N_{W} = P \expect{|h_W|^2}/N_{W} = \overline \gamma_W.\]
Since the channel fading coefficients $h$ are zero-mean complex
Gaussian random variables and the instantaneous SNR $\gamma \propto
|h|^2$, it follows that $\gamma$ is exponentially distributed,
specifically
\begin{equation}
p(\gamma_M)=\frac{1}{\overline
\gamma_M}\exp\left({-\frac{\gamma_M}{\overline
\gamma_M}}\right),\qquad \gamma_M>0 \label{eq:pdf1}
\end{equation}
and
\begin{equation}
p(\gamma_W)=\frac{1}{\overline
\gamma_W}\exp\left({-\frac{\gamma_W}{\overline
\gamma_W}}\right),\qquad \gamma_W>0. \label{eq:pdf2}
\end{equation}

\subsection{Problem Statement}

Let the transmission rate between Alice and Bob be $R=H(W^k)/n$, the
equivocation rate\footnote{Notice that the secrecy condition used here (and in~\cite{Wyner:BSTJ75, CheongH:78}) is weaker than the one proposed by Maurer and Wolf in~\cite{MaurerW:00}, where the information obtained by the eavesdropper is negligibly small not just in terms of rate but in absolute terms. Unfortunately, it is unclear whether the
techniques used for discrete memoryless channels in~\cite{MaurerW:00} can be
extended for Gaussian channels, in particular information reconciliation and
privacy amplification. Resolving this issue is part of our ongoing efforts.}
or Eve's uncertainty be
$\Delta=H(W^k|Y^n_W)/H(W^k)$, and the error probability
$\mathcal{P}_\epsilon = \mathcal{P} (W^k \neq \hat W^k)$, where
$W^k$ denotes the sent messages and $\hat W^k$ denotes Bob's
estimate of the sent messages.

In general, one is interested in characterizing the
rate-equivocation region, that is, the set of achievable pairs
$(R',d')$. A pair $(R',d')$ is achievable if for all $\epsilon>0$
there exists an encoder-decoder pair such that $R \geq R'-\epsilon$,
$\Delta\geq d'-\epsilon$, and $\mathcal{P}_\epsilon \leq
\epsilon$. Here, however, we are interested in characterizing the
{\it secrecy capacity} $C_s$, that is, the maximum transmission rate
$R$ at $\Delta=1$.

In the rest of the paper, we will study the secrecy capacity of this
wireless system for different channel state information (CSI)
regimes. We will always assume that Bob has perfect knowledge of the
main channel fading coefficient and that Eve also has perfect
knowledge of the wiretap channel fading coefficient \footnote{By
virtue of the independence of the main channel and the wiretap
channel, there are no additional benefits/penalities if Bob knows
the wiretap fading coefficient and/or Eve knows the main channel
fading coefficient}. We will also always assume that Alice has
perfect knowledge of the main channel fading coefficient. Note that
these assumptions are realistic for this slow fading wireless
environment: both receivers can always obtain close to perfect
channel estimates and, additionally, the legitimate receiver can
also feedback the channel estimates to the legitimate transmitter.
However, we will assume various regimes for Alice knowledge of the
eavesdropper channel:
\begin{enumerate}
\item No knowledge of the wiretap channel fading coefficient;

\item Partial knowledge of the wiretap channel fading coefficient;

\item Perfect knowledge of the wiretap channel fading coefficient.
\end{enumerate}
Case 1 corresponds to the situation where Eve is a passive and  malicious
eavesdropper in the wireless network. Cases 2 and 3 correspond to
the situation where Eve is another active user in the wireless
network, so that, \eg in a TDMA environment, Alice can estimate the
wiretap channel during Eve transmissions.

In the following sections, we will characterize the secrecy capacity
in terms of outage events for the wireless system setup in
\fig{fig:wireless_system_setup}.

\section{Secrecy Capacity and Outage without CSI on the Eavesdropper's Channel}
\label{sec:outage}

In this section, we will consider the situation where the legitimate
transmitter (Alice) knows nothing about the state of the
eavesdropper's channel. However, we assume that the legitimate
transmitter and receiver know the state of the main channel
perfectly and that the eavesdropper also knows the state of the
eavesdropper channel perfectly (see \secref{sec:statement}).

Consequently, this section characterizes the secrecy capacity of a
quasi-static Rayleigh fading channel in terms of outage probability.
First, we consider a single realization of the fading coefficients
and compute its instantaneous secrecy capacity. Then, we discuss the
existence of (strictly positive) secrecy capacity in the general
case, and build upon the resulting insights to characterize the
outage probability and the outage secrecy capacity.

\subsection{Instantaneous Secrecy Capacity}
\label{sec:preliminaries}
We start by deriving the secrecy capacity for one realization of a
pair of quasi-static fading channels with complex noise and
complex fading coefficients.

For this purpose, we recall the results of~\cite{CheongH:78} for
 the real-valued Gaussian wiretap channel, where it is assumed that Alice and
Bob communicate over a standard real additive white Gaussian noise
(AWGN) channel with noise power $N_M$ and Eve's observation is also
corrupted by Gaussian noise with
 power $N_W>N_M$, i.e.~Eve's receiver has lower SNR than
 Bob's. The power is constrained according to
 $\frac{1}{n}\sum_{i=1}^n\expect{X(i)^2}\leq P$. For this instance, the
 secrecy capacity is given by
 \begin{equation}
 C_s=C_M-C_W,
\label{eq:diff}
 \end{equation}
 where
 \[C_M
 =\frac{1}{2}\log\left(1+\frac{P}{N_M}\right)\]
 is the capacity of the main channel and
 \[C_W=\frac{1}{2}\log\left(1+\frac{P}{N_W}\right)\]
 denotes the capacity\footnote{Unless otherwise specified, all logarithms
 are taken to base two.}
 of the eavesdropper's channel.
From this result, we can derive the following lemma which describes the instantaneous secrecy capacity for the wireless fading scenario defined in \secref{sec:statement}.

\begin{lemma}
The secrecy capacity for one realization of the quasi-static complex fading
wiretap-channel is given by
 \begin{equation}
C_s = \left\{ \begin{array}{ll} \log\left(1+\gamma_M\right)-
\log\left(1+\gamma_W\right)& \textrm{if $\gamma_M > \gamma_W$}\\
0 & \textrm{if $\gamma_M \leq \gamma_W$.}
\end{array} \right.
\label{eq:cs}
\end{equation}
\end{lemma}\vspace{0.2cm}
\begin{proof}
Suppose that both the main and the wiretap channel are complex
AWGN channels, i.e.~ transmit and receive symbols are complex and
both additive noise processes are zero mean circularly symmetric
complex Gaussian. The power of the complex input $X$ is constrained
according to $\frac{1}{n}\sum_{i=1}^n\expect{|X(i)|^2}\leq P$. Since
each use of the complex AWGN channel can be viewed as two uses of a
real-valued AWGN channel~\cite[Appendix B]{Tse:book}, the secrecy
capacity of the complex wiretap channel follows from \eq{eq:diff} as
 \[C_s=\log\left(1+\frac{P}{N_M}\right)-
\log\left(1+\frac{P}{N_W}\right),\] per complex
dimension\footnote{Alternatively, this result can be proven by
repeating step by step the proofs of~\cite{CheongH:78} using
complex-valued random variables instead of real-valued ones.}.

To complete the proof, we introduce complex fading coefficients for both
the main channel and the eavesdropper's channel, as detailed in
\secref{sec:statement}. Since in the quasi-static case $h_M$ and
$h_W$ are random but remain constant for all time,  it is perfectly
reasonable to view the main channel (with fading) as a complex AWGN
channel~\cite[Chapter 5]{Tse:book} with SNR
$\gamma_M=P|h_M|^2/N_{M}$ and capacity
\[C_M
 =\log\left(1+|h_M|^2\frac{P}{N_M}\right).\]
Similarly, the capacity of the eavesdropper's channel
is given by
\[C_W=\log\left(1+|h_W|^2\frac{P}{N_W}\right),\]
with SNR $\gamma_W=P|h_W|^2/N_{W}$. Thus, once again based on
\eq{eq:diff} and the nonnegativity of channel capacity, we may
write the secrecy capacity for one realization of
 the quasi-static fading scenario as \eq{eq:cs}.
\end{proof}

\subsection{Probability of Strictly Positive Secrecy Capacity}

We will now determine the probability $\prob(C_s > 0)$ of a strictly positive
secrecy capacity between Alice and Bob.
\begin{lemma}
For average signal-to-noise ratios $\overline{\gamma}_M$ and $\overline{\gamma}_W$ on the main channel and the wiretap channel, respectively, we have that
\begin{eqnarray}
\prob(C_s > 0) &=&
\frac{\overline{\gamma}_M}{\overline{\gamma}_M+\overline{\gamma}_W}.
\label{eq:prob}
\end{eqnarray}
\end{lemma}\vspace{0.2cm}
\begin{proof}
As explained in \secref{sec:preliminaries}, for
specific fading realizations, the main channel (from Alice to Bob)
and the eavesdropper's channel (from Alice to Eve) can be viewed as
complex AWGN channels with SNR $\gamma_M$ and $\gamma_W$,
respectively. Moreover, from \eq{eq:cs} it follows that the secrecy
capacity is positive when $\gamma_M > \gamma_W$ and is zero when
$\gamma_M \leq \gamma_W$. Invoking independence between the main
channel and the eavesdropper's channel and knowing that the random
variables $\gamma_M$ and $\gamma_W$ are exponentially distributed
with probability density functions given by \eq{eq:pdf1} and
\eq{eq:pdf2}, respectively, we may write the probability of
existence of a non-zero secrecy capacity as
\begin{eqnarray}
\prob(C_s > 0) &=& \prob(\gamma_M > \gamma_W)\nonumber\\
&=& \int_{0}^{\infty} \int_{0}^{\gamma_M}
p(\gamma_M,\gamma_W) d \gamma_W d \gamma_M \nonumber\\
 &=& \int_{0}^{\infty} \int_{0}^{\gamma_M}
p(\gamma_M) p(\gamma_W) d \gamma_W d \gamma_M \nonumber\\
&=&
\frac{\overline{\gamma}_M}{\overline{\gamma}_M+\overline{\gamma}_W}.
\nonumber
\end{eqnarray}
\end{proof}

It is also useful to express this probability in terms of parameters
related to user location.
\begin{cor}
For distance $d_M$ between Alice and Bob, distance $d_W$ between Alice and Eve, and pathloss exponent $\alpha$, we have that
\begin{equation}
\prob(C_s > 0) = \frac{1}{1+(d_M/d_W)^\alpha}
\label{eq:prob1}
\end{equation}
\end{cor}
\begin{proof}
The corollary follows directly from the fact that $\overline{\gamma}_M
\varpropto 1/d_M^{\alpha}$ and $\overline{\gamma}_W \varpropto
1/d_W^{\alpha}$~\cite{Rappaport:book}.
\end{proof}
\begin{remark}
Note that when $\gamma_M \gg \gamma_W$ (or $d_M \ll d_W$) then
$\prob(C_s > 0) \approx 1$ (or $\prob(C_s = 0) \approx 0$).
Conversely, when $\gamma_W \gg \gamma_M$ (or $d_W \ll d_M$) then
$\prob(C_s > 0) \approx 0$ (or $\prob(C_s = 0) \approx 1$).
It is also interesting to observe that to guarantee the existence of a
non-zero secrecy capacity with probability greater than $p_0$ then
it follows from \eq{eq:prob} and \eq{eq:prob1} that
\[
\frac{\overline{\gamma}_M}{\overline{\gamma}_W} > \frac{p_0}{1-p_0}
\]
or
\[
\frac{d_M}{d_W} < \sqrt[\alpha]{\frac{1-p_0}{p_0}}.
\]
In particular, a non-zero secrecy capacity exists even when
$\overline{\gamma}_M < \overline{\gamma}_W$ or $d_M
> d_W$, albeit with probability less than $1/2$.
\end{remark}

\subsection{Outage Probability of Secrecy Capacity}

We are now ready to characterize the outage probability
\[
\pout(R_s) = \prob(C_s < R_s),
\] i.e.~the
probability that the instantaneous secrecy capacity is less than a
target secrecy rate $R_s>0$. The operational significance of this
definition of outage probability is that when setting the secrecy
rate $R_s$ Alice is assuming that the capacity of the wiretap
channel is given by $C'_W=C_M-R_s$. As long as $R_s<C_s$, Eve's
channel will be worse than Alice's estimate, i.e. $C_W<C_W'$, and
so the wiretap codes used by Alice will ensure perfect secrecy.
Otherwise, if $R_s>C_s$ then $C_W>C_W'$ and information-theoretic
security is compromised.
\begin{theorem}
The outage probability for a target secrecy rate $R_s$ is given by
\begin{equation}
\pout(R_s) = 1 - \frac{\overline{\gamma}_M}{\overline{\gamma}_M +
2^{R_s} \overline{\gamma}_W}
\exp\left(-\frac{2^{R_s}-1}{\overline{\gamma}_M}\right).
\label{eq:pout}
\end{equation}
\end{theorem}\vspace{0.2cm}
\begin{proof}
Invoking the total probability theorem,
\begin{eqnarray*}
\pout(R_s) &=& \prob(C_s < R_s \mid \gamma_M >
\gamma_W)\prob(\gamma_M > \gamma_W)\\&& + \prob(C_s < R_s \mid
\gamma_M \leq \gamma_W)\prob(\gamma_M \leq \gamma_W)
\end{eqnarray*}
Now, from \eq{eq:prob} we know that
\[
\prob(\gamma_M > \gamma_W) =
\frac{\overline{\gamma}_M}{\overline{\gamma}_M+\overline{\gamma}_W}.
\]
Consequently, we have
\[
\prob(\gamma_M \leq \gamma_W) = 1 - \prob(\gamma_M > \gamma_W) =
\frac{\overline{\gamma}_W}{\overline{\gamma}_M+\overline{\gamma}_W}.
\]
On the other hand, we also have that
\[
\prob(C_s < R_s \mid \gamma_M > \gamma_W) \qquad\hspace{6cm}
\]
\begin{eqnarray*}
&=&\prob(\log(1+\gamma_M)-\log(1+\gamma_W) < R_s \mid \gamma_M >
\gamma_W)\\
&=& \prob (\gamma_M < 2^{R_s} (1+\gamma_W) - 1 \mid \gamma_M >
\gamma_W)\\
&=& \int_{0}^{\infty} \int_{\gamma_W}^{2^{R_s}(1+\gamma_W)-1}
p(\gamma_M,\gamma_W \mid \gamma_M > \gamma_W) d \gamma_W d \gamma_M\\
&=&  \int_{0}^{\infty} \int_{\gamma_W}^{2^{R_s}(1+\gamma_W)-1}
\frac{p(\gamma_M) p(\gamma_W)}{\prob(\gamma_M > \gamma_W)} d \gamma_W d \gamma_M\\
&=& 1 - \frac{\overline{\gamma}_M+\overline{\gamma}_W}
{\overline{\gamma}_M + 2^{R_s}\overline{\gamma}_W}
\exp\left(-\frac{2^{R_s}-1}{\overline{\gamma}_M}\right)
\end{eqnarray*}
and, since $R_s>0$,
\[
\prob(C_s < R_s \mid \gamma_M \leq \gamma_W) = 1.
\]
Combining the previous five equations, we get
\begin{equation}
\pout(R_s) = 1 - \frac{\overline{\gamma}_M}{\overline{\gamma}_M +
2^{R_s} \overline{\gamma}_W}
\exp\left(-\frac{2^{R_s}-1}{\overline{\gamma}_M}\right).
\end{equation}
\end{proof}

\subsection{Outage Secrecy Capacity}

Another performance measure of interest is the $\epsilon$-outage
secrecy capacity,
 defined as the largest secrecy rate such that
the outage probability is less than $\epsilon$, i.e.
\[
\pout(\cout(\epsilon)) = \epsilon~.
\]
Although it is hard to obtain the outage secrecy capacity
analytically --- the outage probability is a complicated function
of the secrecy rate --- it is possible to compute its value
numerically based on \eq{eq:pout}.

\subsection{Asymptotic Behavior}

It is illustrative to examine the asymptotic behavior of the
outage probability for extreme values of the target secrecy rate
$R_s$. From \eq{eq:pout} it follows that when $R_s \rightarrow 0$,
\[
\pout \rightarrow
\frac{\overline{\gamma}_W}{\overline{\gamma}_M+\overline{\gamma}_W}
\]
and when $R_s \rightarrow \infty$, we have that $ \pout
\rightarrow 1$, such that it becomes impossible for Alice and Bob to
transmit secret information (at very high rates).

\begin{figure}[t]
  \centering
  \includegraphics[width=9cm]{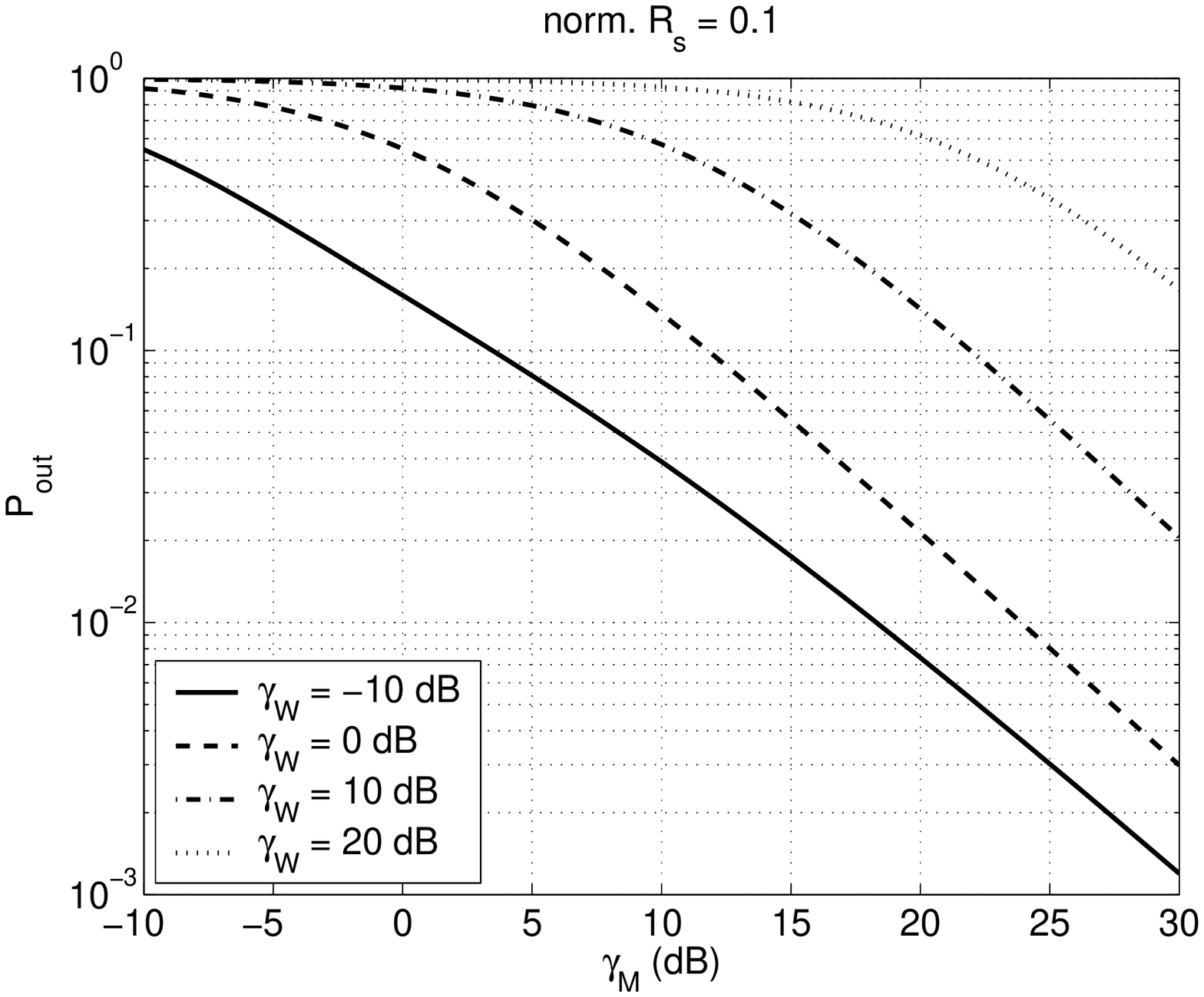}
 \vspace{-0.1cm}
  \caption{Outage probability versus $\overline{\gamma}_M$, for selected values of $\overline{\gamma}_W$
and for a normalized target secrecy rate equal to 0.1.
Normalization is effected with respect to the capacity of an AWGN
channel with SNR equal to $\overline{\gamma}_M$.}
  \label{fig:outage_probability1}
\end{figure}

\begin{figure}[b]
\vspace{-0.5cm}
  \centering
  \includegraphics[width=9cm]{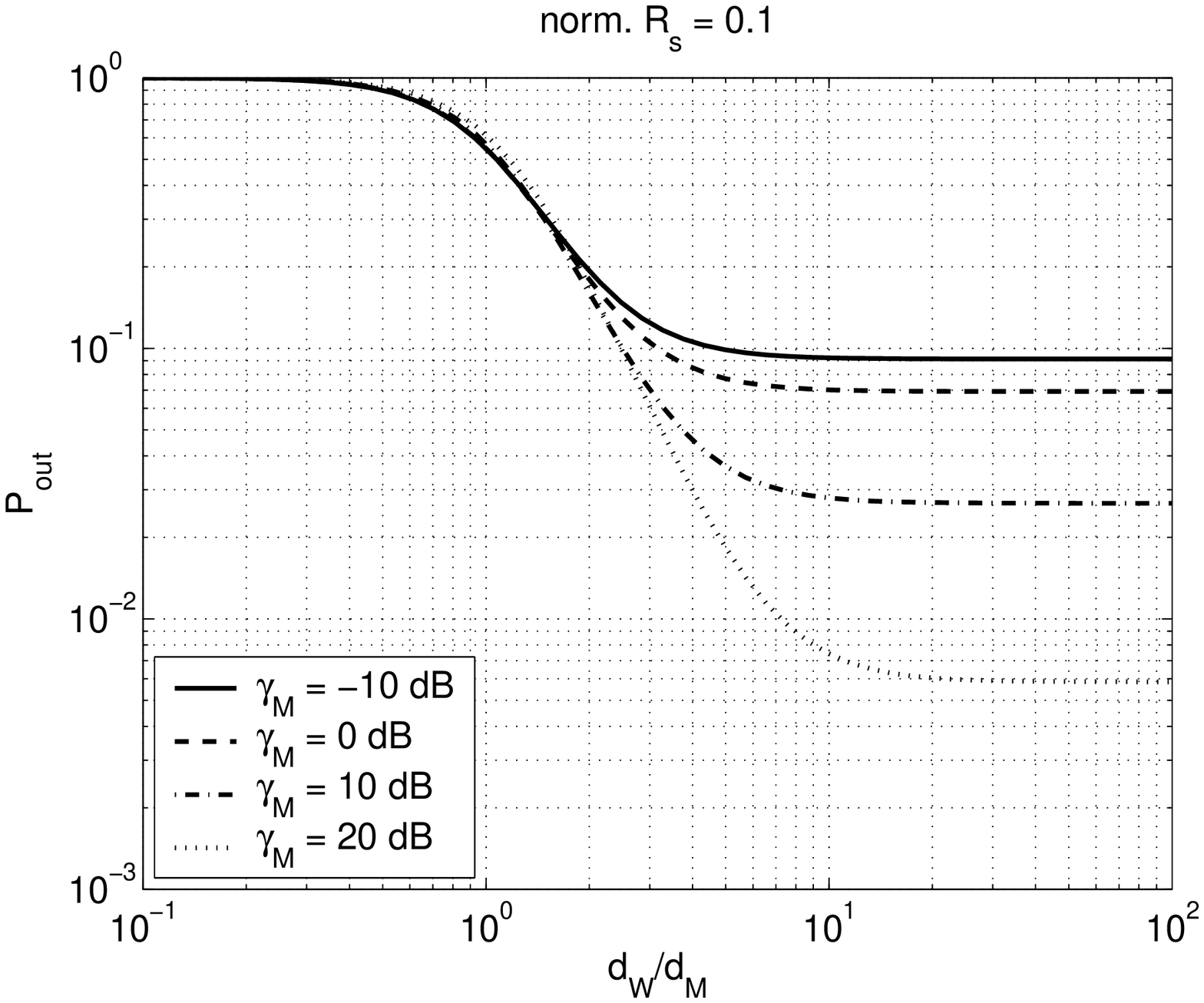}
  \caption{Outage probability versus $d_W/d_M$, for selected values of $\overline{\gamma}_M$
and for a normalized target secrecy rate equal to 0.1.
Normalization is effected with respect to the capacity of an AWGN
channel with SNR equal to $\overline{\gamma}_M$.}
  \label{fig:outage_probability2}

\end{figure}

Also of interest is the asymptotic behavior of the outage
probability for extreme values of the average SNRs of the main
channel and the eavesdropper's channel. When $\overline{\gamma}_M
\gg \overline{\gamma}_W$, equation \eq{eq:pout} yields
\[
\pout (R_s) \approx 1 - \exp
\left(-\frac{2^{R_s}-1}{\overline{\gamma}_M}\right),
\]
and in a high SNR regime $ \pout \approx
(2^{R_s}-1)/\overline{\gamma}_M$, i.e.~the outage probability
decays as $1/\overline{\gamma}_M$. Conversely, when
$\overline{\gamma}_W \gg \overline{\gamma}_M$,
\[
\pout (R_s) \approx 1,
\]
and confidential communication becomes impossible.

\fig{fig:outage_probability1} depicts the outage probability
versus $\overline{\gamma}_M$, for selected values of
$\overline{\gamma}_W$ and for a normalized target secrecy rate
equal to 0.1. Observe that the higher $\overline{\gamma}_M$ the
lower the outage probability, and the higher $\overline{\gamma}_W$
the higher the probability of an outage. Moreover, if
$\overline{\gamma}_M \gg \overline{\gamma}_W$, the outage
probability decays as $1/\overline{\gamma}_M$. Conversely, if
$\overline{\gamma}_W \gg \overline{\gamma}_M$ the outage
probability approaches one.

With respect to the asymptotic behavior of the outage secrecy
capacity, it is not difficult to see that $\cout \rightarrow 0$
yields $\pout
\rightarrow\overline{\gamma}_W/(\overline{\gamma}_M+\overline{\gamma}_W)$,
and when $\cout \rightarrow \infty$, we have $\pout \rightarrow
1$.

The impact of the distance ratio on the performance is
illustrated in \fig{fig:outage_probability2}, which depicts the
outage probability versus $d_W/d_M$, for selected values of
$\overline{\gamma}_M$ and for a normalized target secrecy rate
equal to 0.1. The pathloss exponent is set to be equal to a
typical value of 3~\cite{Rappaport:book}. When $d_W/d_M
\rightarrow \infty$ (or $\overline{\gamma}_M/\overline{\gamma}_W
\rightarrow \infty$), we have that $\pout \rightarrow 1 -
\exp(-(2^{R_s}-1)/\overline{\gamma}_M)$. If $d_W/d_M \rightarrow
0$ (or $\overline{\gamma}_M/\overline{\gamma}_W \rightarrow 0$),
then $\pout \rightarrow 1$.

\subsection{Fading Channels versus Gaussian Channels}

It is important to emphasize that under a fading scenario --- in
contrast with the Gaussian wiretap channel~\cite{CheongH:78}---
the goal of a strictly positive (outage) secrecy capacity does
{\it not} require the average SNR of the main channel to be
greater than the average SNR of the eavesdropper's channel. This
is due to the fact that in the presence of fading there is always
a finite probability, however small, that the instantaneous SNR of
the main channel $\gamma_M$ is higher than the instantaneous SNR
of the eavesdropper's channel $\gamma_W$.

Specifically, the results in \secref{sec:outage} demonstrate that
a non-zero outage secrecy capacity requires $\overline{\gamma}_M >
\overline{\gamma}_W$ for $\pout < 0.5$, but we may have
$\overline{\gamma}_M < \overline{\gamma}_W$ for $\pout
> 0.5$. In other words, if we are willing to tolerate some
outage, then there is no obstacle to information-theoretic
security over wireless fading channels. In fact, it is possible to
trade off outage probability for outage secrecy capacity: a higher
outage secrecy capacity corresponds to a higher outage
probability, and vice versa.

\begin{figure}[h]
  \centering
  \vspace{-0.08cm}
  \includegraphics[width=9cm]{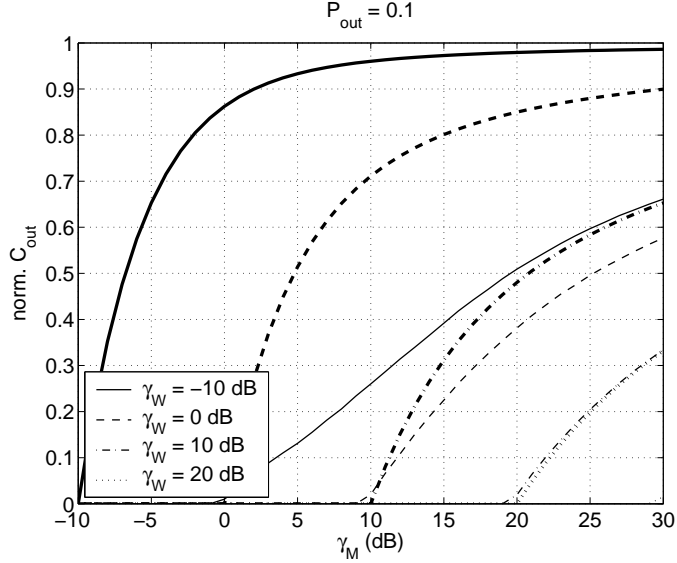}
  \vspace{-0.1cm}
  \caption{Normalized outage secrecy capacity versus $\overline{\gamma}_M$, for selected values of
$\overline{\gamma}_W$, and for an outage probability of 0.1.
Thinner lines correspond to the normalized outage secrecy capacity
in the case of Rayleigh fading channels, while thicker lines
correspond to the secrecy capacity of the Gaussian wiretap
channel. Normalization is effected with respect to the capacity of
an AWGN channel with SNR equal to $\overline{\gamma}_M$.}
  \label{fig:outage_secrecy_capacity1}
\vspace{-0.3cm}
\end{figure}

\begin{figure}[h]
  \centering
  \includegraphics[width=9cm]{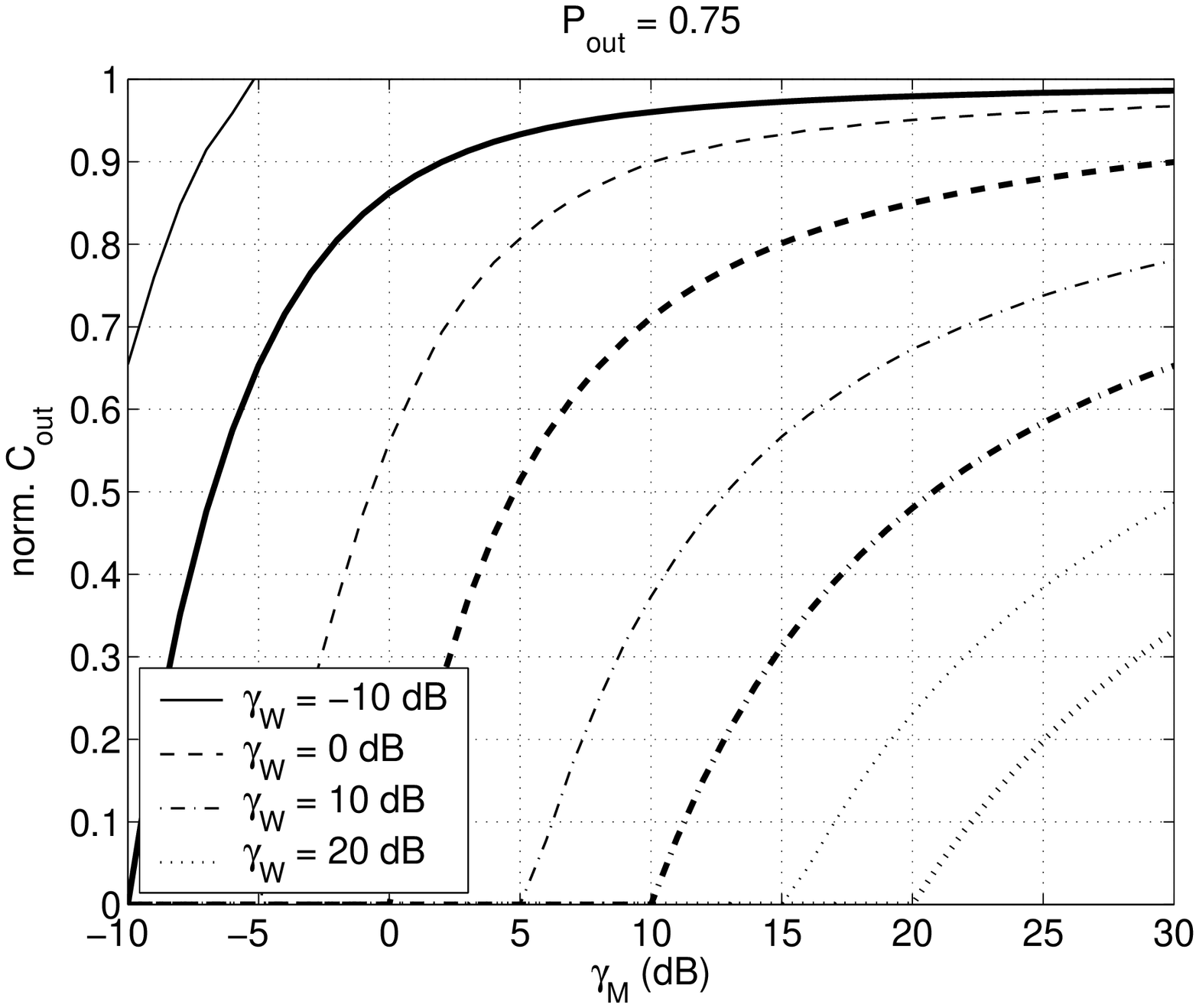}
\vspace{-0.3cm}
  \caption{Normalized outage secrecy capacity versus $\overline{\gamma}_M$, for selected values of
$\overline{\gamma}_W$, and for an outage probability of 0.75.
Thinner lines correspond to the normalized outage secrecy capacity
in the case of the  Rayleigh fading channels, while thicker lines
correspond to the  secrecy capacity of the Gaussian wiretap
channel. Normalization is effected with respect to the capacity of
an AWGN channel with SNR equal to $\overline{\gamma}_M$.}
  \label{fig:outage_secrecy_capacity1_}
\end{figure}

\begin{figure}[h]
\vspace{0.16cm}
  \centering
  \includegraphics[width=9cm]{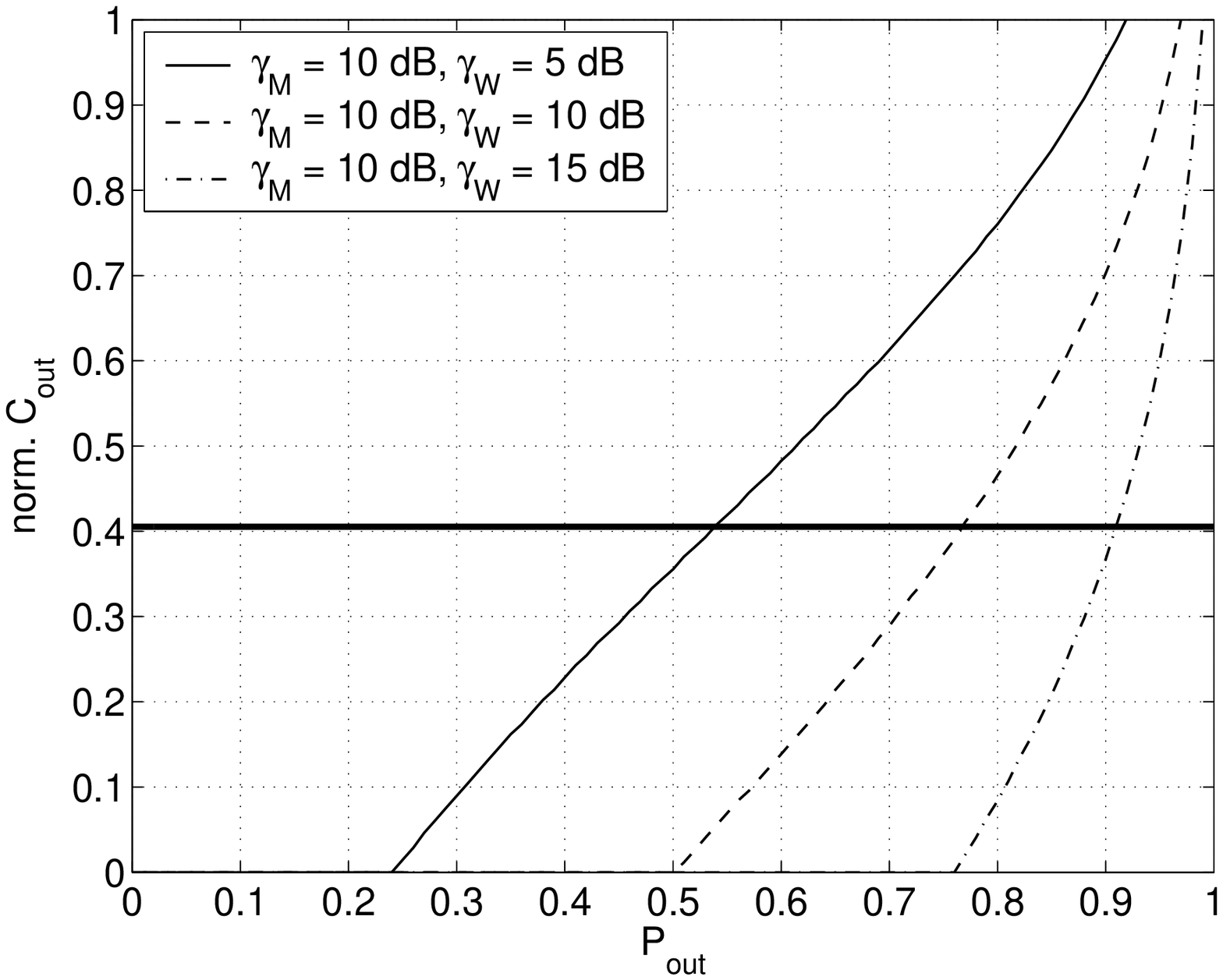}
  \caption{Normalized outage secrecy capacity versus outage probability, for selected values of
$\overline{\gamma}_M$ and $\overline{\gamma}_W$ Thinner lines
correspond to the normalized outage secrecy capacity of the
eavesdropper's Rayleigh fading channel, while thicker lines
correspond to the  secrecy capacity of the Gaussian wiretap
channel (in the last two cases this capacity is zero).
Normalization is effected with respect to the capacity of an AWGN
channel with SNR equal to $\overline{\gamma}_M$.}
  \label{fig:outage_secrecy_capacity_vs_outage_probability}
\vspace{-0.5cm}
\end{figure}

It also turns out that the outage secrecy capacity of a fading
channel can actually be higher than the secrecy capacity of a
Gaussian wiretap channel. Consider the examples shown in
\fig{fig:outage_secrecy_capacity1} and
\fig{fig:outage_secrecy_capacity1_}, which depict the normalized
outage secrecy capacity versus $\overline{\gamma}_M$, for selected
values of $\overline{\gamma}_W$, and for an outage probability of
0.1 and 0.75, respectively. The normalized secrecy capacity of the
Gaussian wiretap channel with main channel SNR equal to
$\overline{\gamma}_M$ and wiretap channel SNR equal to
$\overline{\gamma}_W$ is also included for comparison. Observe
that in the Gaussian case the secrecy capacity is zero when
$\overline{\gamma}_M \leq \overline{\gamma}_W$. In contrast, in
the case of Rayleigh fading channels the outage secrecy capacity
is non-zero even when $\overline{\gamma}_M \leq
\overline{\gamma}_W$ (as long as $\pout > 0.5$). More importantly,
the outage secrecy capacity in the Rayleigh fading case exceeds
the secrecy capacity of the equivalent Gaussian wiretap channel,
for higher outage probabilities. These key observations are also
corroborated by
\fig{fig:outage_secrecy_capacity_vs_outage_probability}, which
compares the normalized (outage) secrecy capacity for  fading
channels to the secrecy capacity of Gaussian channels, for various
outage probabilities. \vspace{-0.2cm}

\begin{figure}[h]
  \centering
  \vspace{0.16cm}
  \includegraphics[width=9cm]{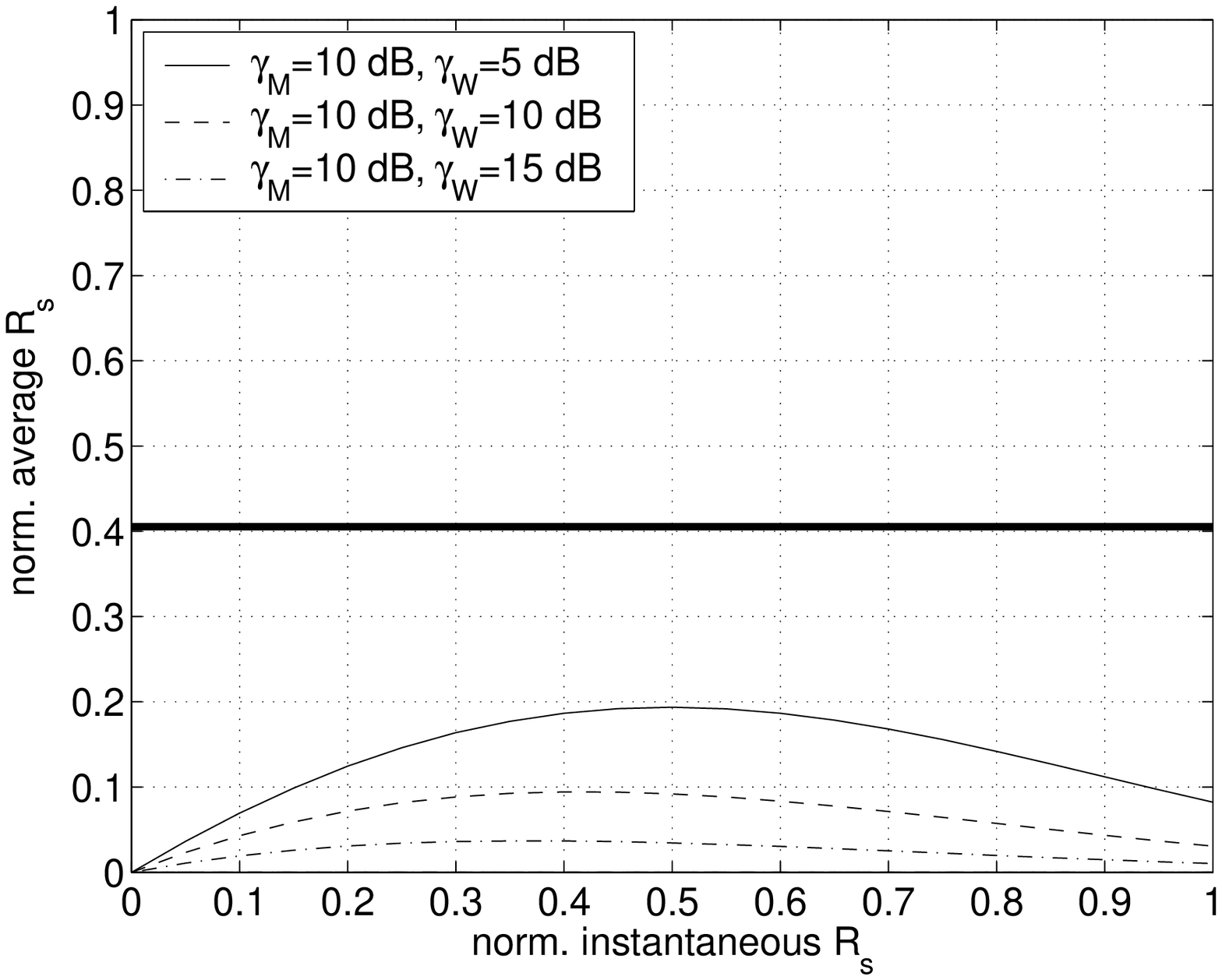}
  \caption{Normalized average secrecy rate versus normalized instantaneous
secrecy rate, for selected values of $\overline{\gamma}_M$ and
$\overline{\gamma}_W$. Thinner lines correspond to the normalized
average secrecy rate of Rayleigh fading channels, while thicker
lines correspond to the secrecy capacity of the Gaussian wiretap
channel (in the last two cases this capacity is zero). Normalization
is effected with respect to the capacity of an AWGN channel with SNR
equal to $\overline{\gamma}_M$.}
  \label{fig:average_secrecy_rate1}
\vspace{-0.5cm}
\end{figure}

\begin{figure}[h]
  \centering
  \vspace{0.16cm}
  \includegraphics[width=9cm]{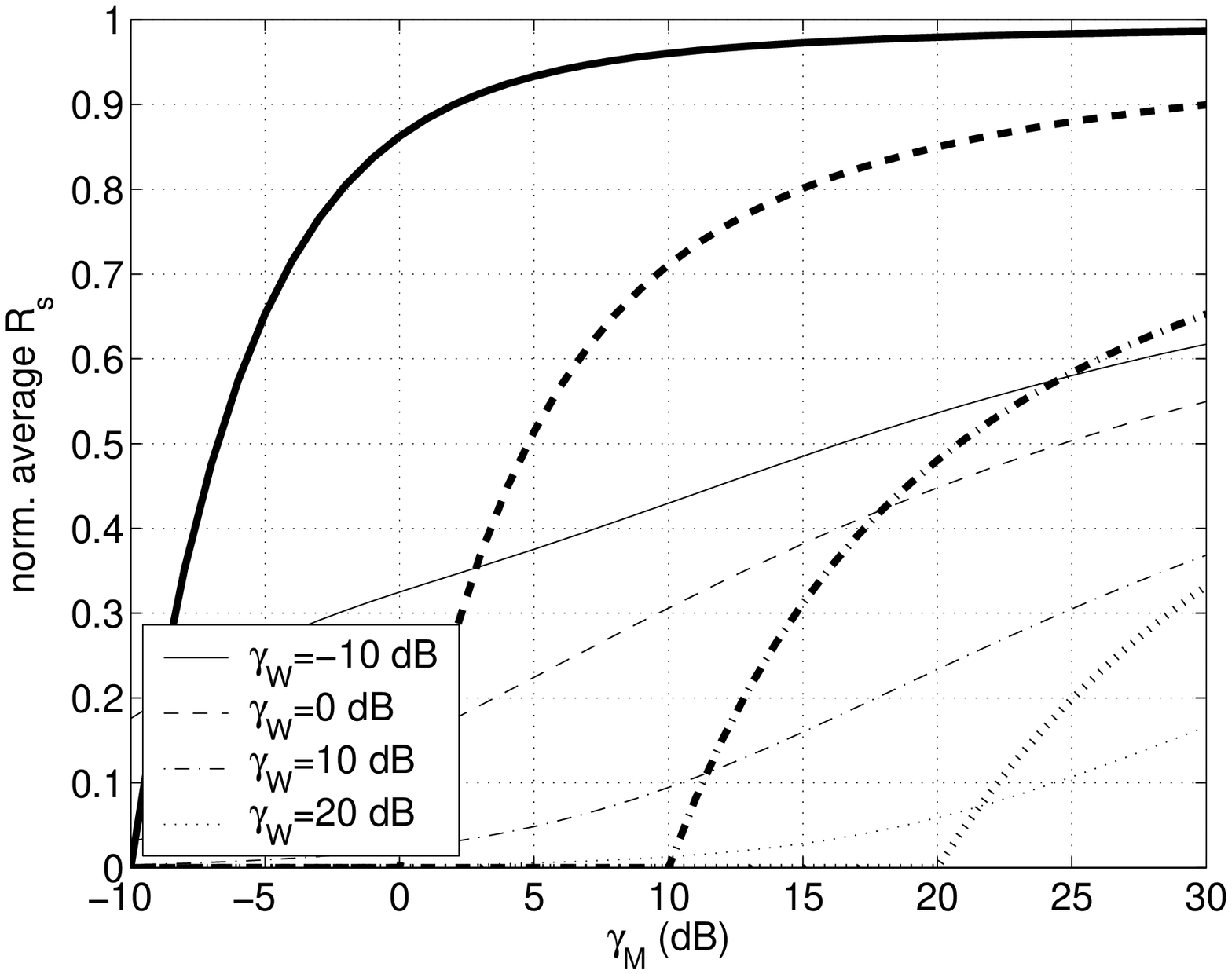}
  \caption{Normalized average secrecy rate versus $\overline{\gamma}_M$, for
selected values of $\overline{\gamma}_W$. Thinner lines correspond
to the normalized average secrecy rate in the case of Rayleigh
fading channels, while thicker lines correspond to the secrecy
capacity of the Gaussian wiretap channel. Normalization is effected
with respect to the capacity of an AWGN channel with SNR equal to
$\overline{\gamma}_M$.}
  \label{fig:average_secrecy_rate2}
\vspace{-0.5cm}
\end{figure}

Finally, it is also interesting to examine the average secrecy rate
given by
\[
\bar{R}_s = (1 - \pout(R_s)) \cdot R_s
\]
The average secrecy rate $\bar{R}_s$ is a function of Alice's target
instantaneous secrecy rate $R_s$, so that Alice is in principle able
to optimize the target instantaneous secrecy rate to maximize the
average secrecy rate (see \fig{fig:average_secrecy_rate1}).
\fig{fig:average_secrecy_rate2} compares the optimum average secrecy
rate in the case of Rayleigh fading channels to the secrecy capacity
of AWGN channels. It is interesting to observe once again that there
is a positive secrecy rate in a Rayleight fading channel even when
the average SNR in the main channel is lower than that in the
eavesdropper channel.

\section{Performance Analysis with Perfect and Imperfect CSI on the Eavesdropper's Channel}
\label{sec:csi}

In this section, we move from the paradigm where the legitimate
transmitter (Alice) knows nothing about the state of the
eavesdropper's channel to one where Alice knows the state of the
eavesdropper's channel partially or even perfectly. However, we
still assume that the legitimate transmitter and receiver know the
state of the main channel perfectly and that the eavesdropper also
knows the state of the eavesdropper channel perfectly (see
\secref{sec:statement}).

We model Alice's estimate of Bob's channel as
\[
\hat{h}_M = h_M,
\]
where $\hat{h}_M$ is the estimate fading coefficient of the main
channel and $h_M$ is the true fading coefficient of the main
channel. Thus, the estimate main channel instantaneous SNR is equal
to the true main channel instantaneous SNR, that is
\[
\hat{\gamma}_M = \gamma_M.
\]
We also model Alice's estimate of Eve's channel as
\[
\hat{h}_W = h_W + \delta_W,
\]
where $\hat{h}_W$ is the estimate fading coefficient of the wiretap
channel, $h_W$ is the true fading coefficient of the wiretap channel
and $\delta_W$ is a circularly symmetric complex Gaussian random
variable with mean zero and variance $\sigma^2$ per dimension. Thus,
the true value and the estimate of wiretap channel instantaneous SNR
may be different, that is
\[
\hat{\gamma}_W \neq \gamma_W.
\]

In this new scenario, we will assume that Alice always sets the
instantaneous information transmission rate $R_s$ to be equal to the
instantaneous secrecy capacity estimate $\hat{C}_s$ of the channel
where
\[
\hat{C}_s = \left\{ \begin{array}{ll} \hat{C}_M - \hat{C}_W & \textrm{if $\hat{C}_M \geq \hat{C}_W$} \\
0 & \textrm{if $\hat{C}_M < \hat{C}_W$}
\end{array} \right.
\]
and $\hat{C}_M = \log(1+\hat{\gamma}_M)$ is the instantaneous main
channel capacity estimate and $\hat{C}_W = \log(1+\hat{\gamma}_W)$
is the instantaneous wiretap channel capacity estimate. We will now
characterize the fundamental secrecy limits when Alice knows the
state of the eavesdropper's channel both imperfectly and perfectly,
including the probability of a secrecy outage, the average secure
throughput (from Alice to Bob) and the average leaked throughput
(from Alice to Eve).

\subsection{Imperfect Knowledge of CSI of Eavesdropper's Channel}

In this situation, Alice conveys information to Bob at a rate $R_s =
\hat{C}_s$ using a wiretap code designed for the operating point
$(\hat{C}_M,\hat{C}_W) = (C_M,\hat{C}_W)$, when $\hat{C}_M
>\hat{C}_W$. If $\hat{C}_W > C_W$ (i.e., $\hat{C}_s < C_s$)
transmission in perfect secrecy is guaranteed, that is, a secrecy
outage does not occur. Otherwise, if $\hat{C}_W < C_W$ (i.e.,
$\hat{C}_s > C_s$) transmission in perfect secrecy cannot be
guaranteed, that is, a secrecy outage occurs. It is now relevant to
characterize the probability of a secrecy outage.

\begin{theorem}
The probability of a secrecy outage is upper bounded by
\begin{equation}
\pout \leq \frac{1}{2} - \frac{1}{2} \frac{1}{\sqrt{1+2/\sigma^2}}.
\end{equation}
\end{theorem}\vspace{0.2cm}

\begin{proof}
The probability of a secrecy outage is given by
\[
\pout = \prob(\hat{C}_W < C_M, \hat{C}_W < C_W) =
\prob(\hat{\gamma}_W < \gamma_M, \hat{\gamma}_W < \gamma_W) =
\prob(\hat{\gamma}_W < min(\gamma_M,\gamma_W))
\]
Consequently, the probability of a secrecy outage is upper bounded
by
\[
\pout = \prob(\hat{\gamma}_W < min(\gamma_M,\gamma_W)) \leq
\prob(\hat{\gamma}_W < \gamma_W)
\]
Now, $\prob(\hat{\gamma}_W < \gamma_W)$ can be written as follows
\[
\prob(\hat{\gamma}_W < \gamma_W) = \int_{0}^{\infty}
\prob(\hat{\gamma}_W < \gamma_W | \gamma_W) p(\gamma_W) d \gamma_W
\]
where $p(\gamma_W)$ is the probability density function of
$\gamma_W$ (see \eq{eq:pdf2}). Moreover, $\prob(\hat{\gamma}_W <
\gamma_W | \gamma_W)$ can also be written as follows
\[
\prob(\hat{\gamma}_W < \gamma_W | \gamma_W) = \int_{0}^{\gamma_W}
p(\hat{\gamma}_W | \gamma_W) d \gamma_W
\]
where $p(\hat{\gamma}_W | \gamma_W)$ is the probability density
function of $\hat{\gamma}_W$ conditioned on $\gamma_W$. This
probability density function is non-central $\chi^2$ with two
degrees of freedom, i.e.
\[
p(\hat{\gamma}_W | \gamma_W) = \frac{1}{2 \bar{\gamma}_W \sigma^2}
e^{-\frac{(\gamma_W+\hat{\gamma}_W)}{2 \bar{\gamma}_W \sigma^2}} I_0
\Bigg(\frac{\sqrt{\gamma_W \hat{\gamma}_W}}{\bar{\gamma}_W
\sigma^2}\Bigg) , \hat{\gamma}_W > 0
\]
where $I_0(\cdot)$ is the zeroth-order modified Bessel function of
the first kind~\cite{Proakis:01}. Thus, the probability
$\prob(\hat{\gamma}_W < \gamma_W | \gamma_W)$ reduces to
\[
\prob(\hat{\gamma}_W < \gamma_W | \gamma_W) = 1 - Q_1
\big(\sqrt{\gamma_W/(\bar{\gamma}_W
\sigma^2)},\sqrt{\gamma_W/(\bar{\gamma}_W \sigma^2)} \big)
\]
where $Q_1(\cdot,\cdot)$ is the generalized Marcum $Q$
function~\cite{Proakis:01}. Moreover, using standard results for
integrals involving the generalized Marcum $Q$
function~\cite{Simon2003}, the upper bound to the outage probability
reduces to
\begin{equation}
\label{eq:Poutage} \pout \leq \frac{1}{2} - \frac{1}{2}
\frac{1}{\sqrt{1+2/\sigma^2}}.
\end{equation}
\end{proof}

It is also relevant to characterize two other quantities with
operational significance: the average secure throughput (or average
secrecy rate) and the average leaked throughput. These quantities
correspond to the average of the instantaneous secure throughput and
the instantaneous leaked throughput over every possible realization
of the main channel and the eavesdropper's channel. Now, the average
secure throughput is lower bounded by the average of the
transmission rate over instances where the secrecy capacity estimate
is lower than the true secrecy capacity, i.e.
\[
\bar{R}_s \geq \int_{0}^{\infty} \hat{C}_s \,p(\hat{C}_s | \hat{C}_s
< C_s) \,d \hat{C}_s
\]
In turn, the average leaked throughput is upper bounded by the
average of the transmission rate over instances where the secrecy
capacity estimate is higher than the true secrecy capacity, i.e.
\[
\bar{R}_l \leq \int_{0}^{\infty} \hat{C}_s \,p(\hat{C}_s | \hat{C}_s
> C_s) \,d \hat{C}_s
\]
These quantities will be characterized numerically due to the
difficulty in determining closed-form expressions.

\begin{figure}[h]
  \centering
  \vspace{0.16cm}
  \includegraphics[width=9cm]{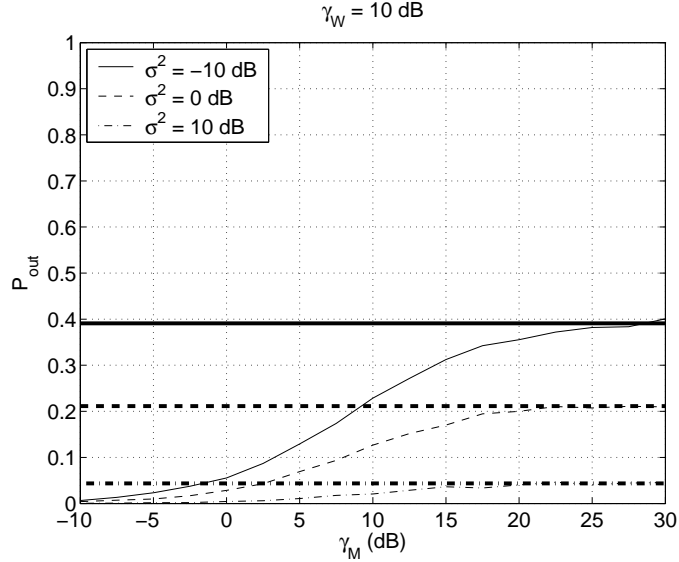}
  \caption{Outage probability versus $\overline{\gamma}_M$ for $\overline{\gamma}_W = 10$ dB, and for selected values of
$\sigma^2$. Thicker lines correspond to the upper bound to the
outage probability while thinner lines correspond to the true outage
probability.}
  \label{fig:outage_probability_imperfect_csi}
\vspace{-0.5cm}
\end{figure}

A number of comments on the behavior of the various performance
measures are now in order.
\fig{fig:outage_probability_imperfect_csi} shows that the upper
bound to the outage probability is considerably tight in a regime
where the average SNR of the main channel is greater than the
average SNR of the eavesdropper channel. More importantly, the
outage probability is a monotone decreasing function of the variance
of the channel estimation error, so that for $\sigma^2
> 0$ the higher the variance of the channel estimation errors the
lower the outage probability.

\begin{figure}[h]
  \centering
  \vspace{0.16cm}
  \includegraphics[width=9cm]{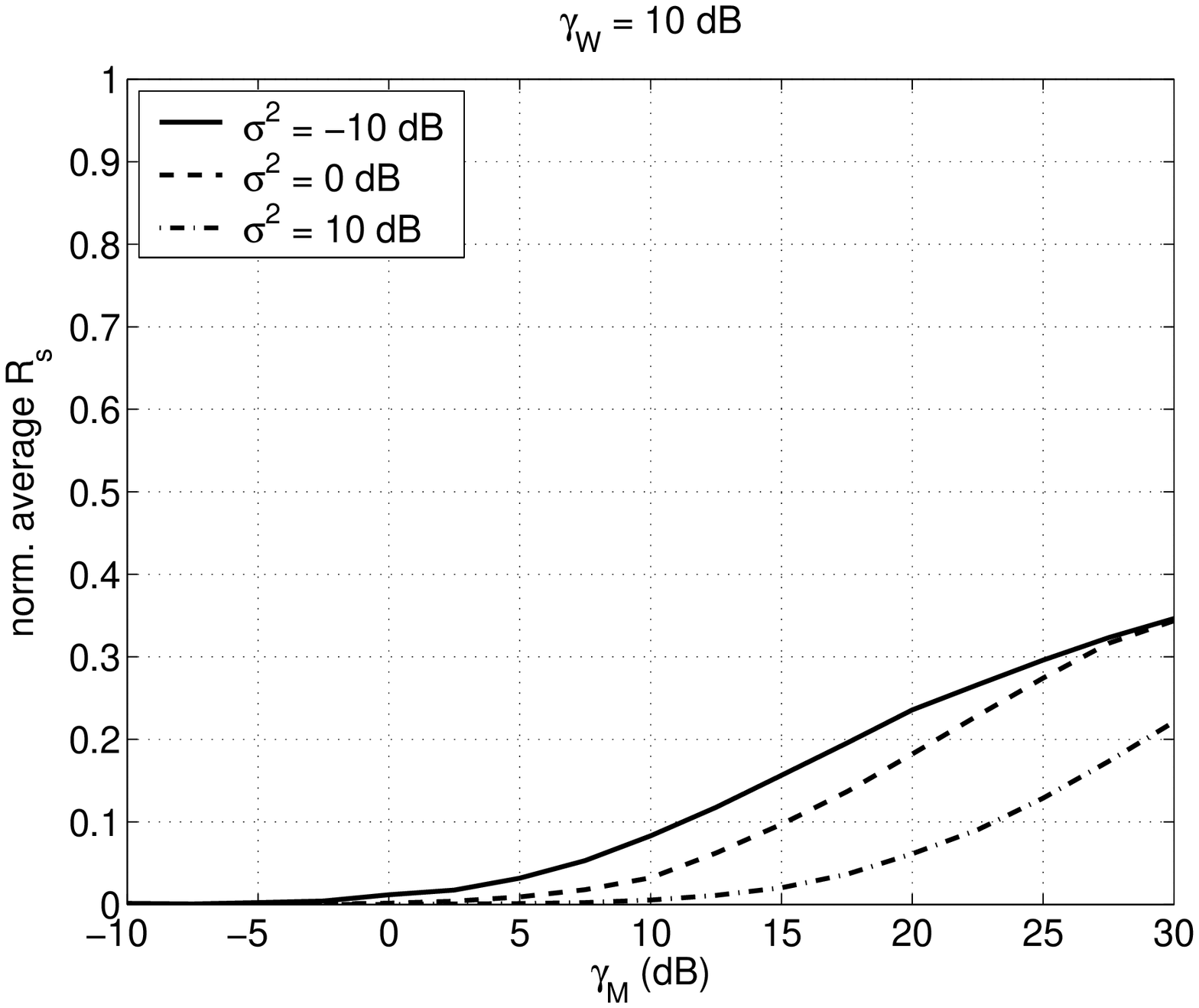}
  \caption{Normalized average secure throughput versus $\overline{\gamma}_M$ for $\overline{\gamma}_W = 10$ dB, and for selected values of
$\sigma^2$. Normalization is effected with respect to the capacity
of an AWGN channel with SNR equal to $\overline{\gamma}_M$.}
  \label{fig:average_secure_throughput}
\vspace{-0.5cm}
\end{figure}

\begin{figure}[h]
  \centering
  \vspace{0.16cm}
  \includegraphics[width=9cm]{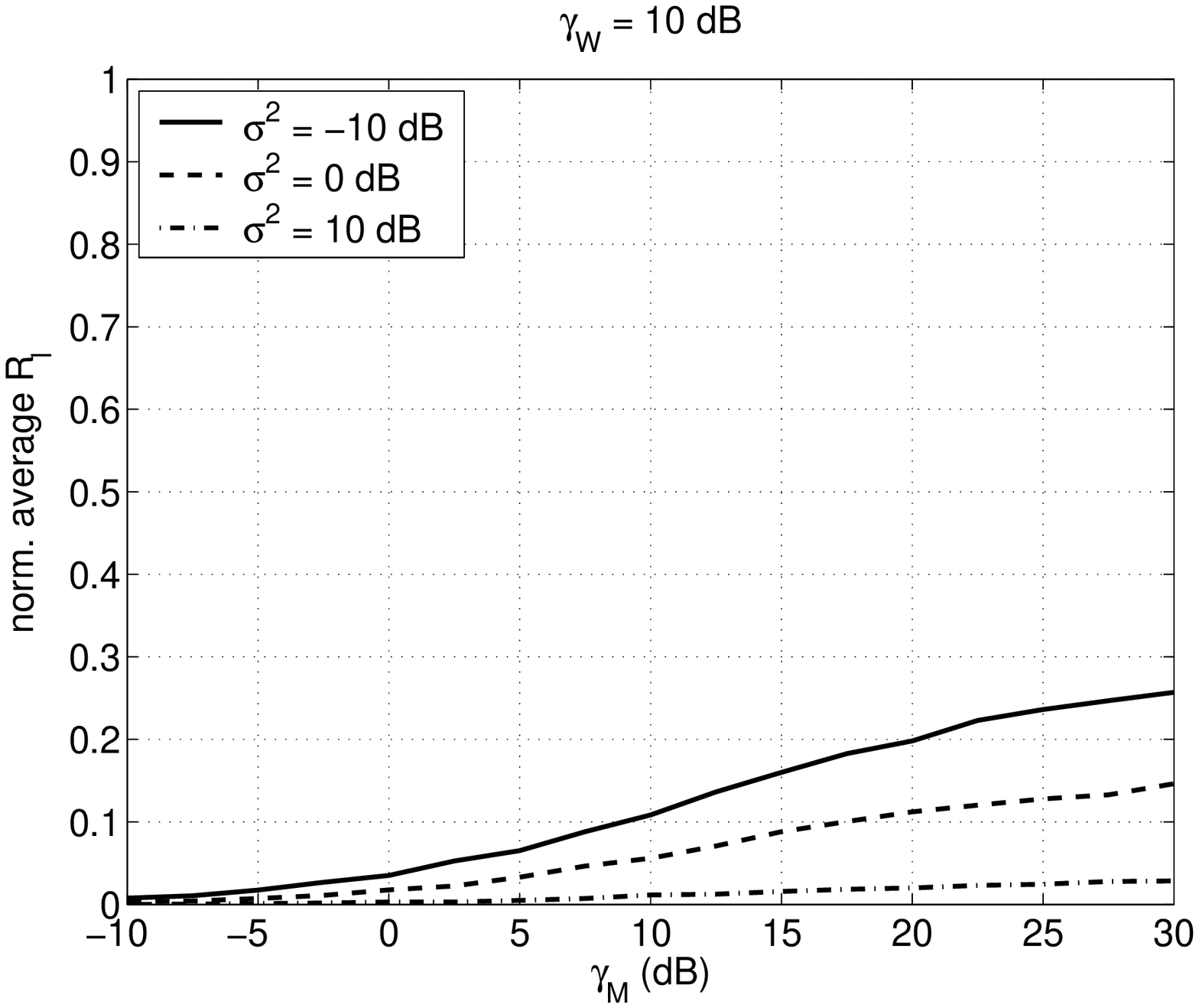}
  \caption{Normalized average leaked throughput versus $\overline{\gamma}_M$ for $\overline{\gamma}_W = 10$ dB, and for selected values of
$\sigma^2$. Normalization is effected with respect to the capacity
of an AWGN channel with SNR equal to $\overline{\gamma}_M$.}
  \label{fig:average_leaked_throughput}
\vspace{-0.5cm}
\end{figure}

This counterintuitive result is based on the fact that for moderate
values of the variance of the channel estimation error Alice tends
to consistently underestimate the secrecy capacity of the system.
Consequently, the attempted instantaneous transmission rate is
consistently lower than the instantaneous secrecy capacity so that
the outage probability is also lower. This in turn results in a
lower average secure throughput and a lower average leaked
throughput as shown in \fig{fig:average_secure_throughput} and
\fig{fig:average_leaked_throughput}.

Yet, of extreme relevance is the fact that even in the presence of
channel estimation errors it is possible to convey information in a
secure manner over a wireless environment (that is, with an average
secure throughput substantially greater than the average leaked
throughput) provided now that the average SNR of the main channel is
greater than the average SNR of the eavesdropper channel (cf.
\fig{fig:average_secure_throughput} and
\fig{fig:average_leaked_throughput}).


\subsection{Perfect Knowledge of CSI of Eavesdropper's Channel}

In this situation, Alice conveys information to Bob at a rate $R_s =
\hat{C}_s = C_s$ using a wiretap code designed for the operating
point $(\hat{C}_M,\hat{C}_W) = (C_M,C_W)$, so that a secrecy outage
never occurs. It follows that the average secure throughput (average
secrecy rate) is
\[
\bar{R}_s = \int_{0}^{\infty} R_s d F_{C_s} (R_s),
\]
where \footnote{Note that the expression for the cumulative
distribution function of the instantaneous secrecy capacity is
exactly the same as the expression for the outage probability in
\eq{eq:pout}.}
\[
F_{C_s}(R_s) = \prob(C_s < R_s) = 1 -
\frac{\bar{\gamma}_M}{\bar{\gamma}_M + 2^{R_s} \bar{\gamma}_W}
e^{\frac{2^{R_s}-1}{\bar{\gamma}_M}},
\]
and the average leaked throughput is zero.

\begin{figure}[h]
  \centering
  \vspace{0.16cm}
  \includegraphics[width=9cm]{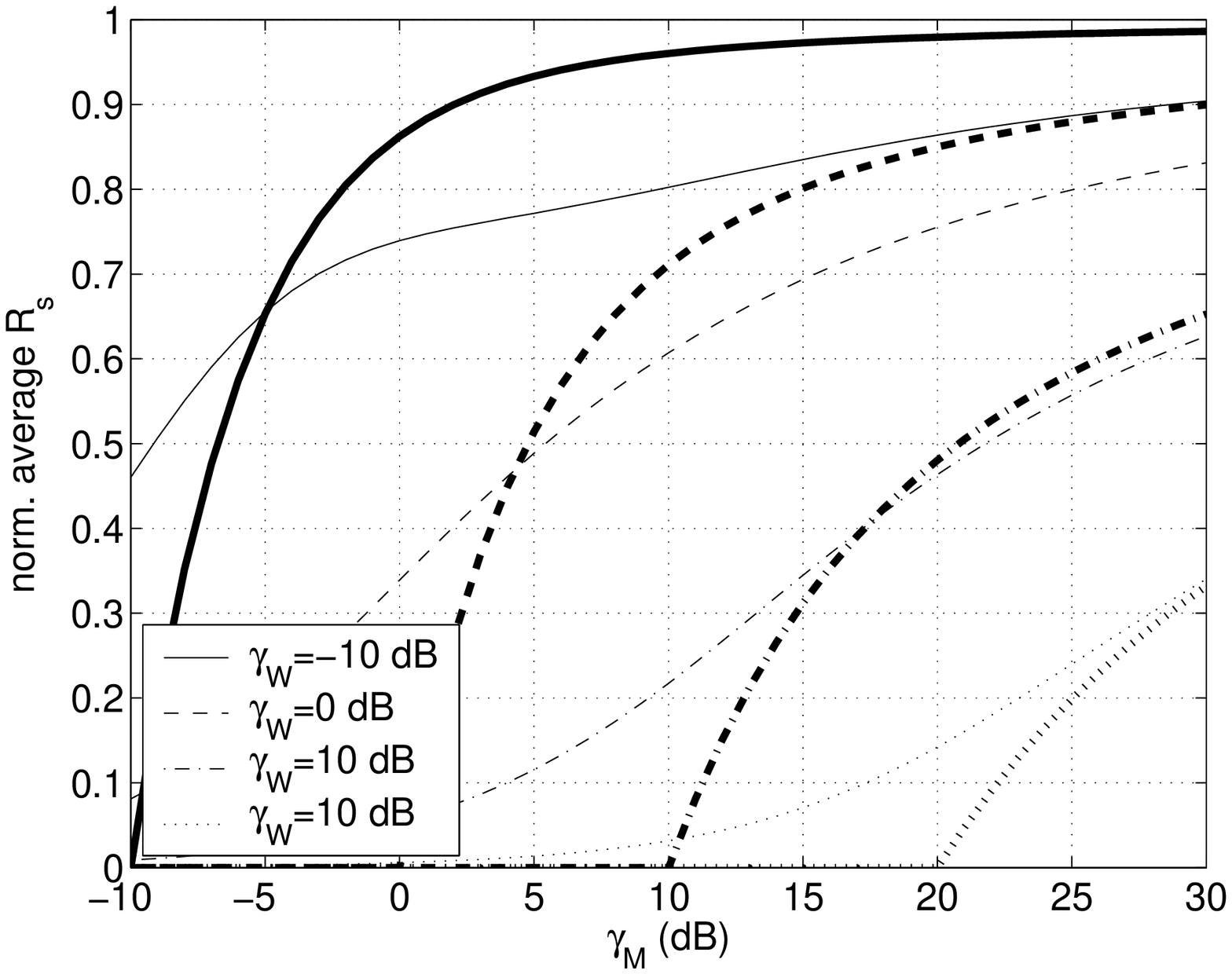}
  \caption{Normalized average secrecy rate versus $\overline{\gamma}_M$, for selected values of
$\overline{\gamma}_W$. Thinner lines correspond to the normalized
average secrecy rate in the case of Rayleigh fading channels, while
thicker lines correspond to the secrecy capacity of the Gaussian
wiretap channel. Normalization is effected with respect to the
capacity of an AWGN channel with SNR equal to
$\overline{\gamma}_M$.}
  \label{fig:average_secrecy_capacity}
\vspace{-0.5cm}
\end{figure}

\fig{fig:average_secrecy_capacity} compares the average secrecy rate
in a "wiretap" Rayleigh fading channel to the secrecy capacity in
the classic wiretap Gaussian channel. Strikingly, one observes that
the average secrecy rate in the fading channel is indeed higher than
or close to the secrecy capacity in the Gaussian channel. One also
observes that, in contrast to the situation in the Gaussian channel,
the average secrecy rate in the fading channel is non-zero even when
the average SNR of the main channel is lower than the average SNR of
the eavesdropper channel. These observations underline once again
the potential of fading channels to secure the transmission of
information between two legitimate parties against a possible
eavesdropper.

\section{Information-Theoretic vs. Computational Security in
Wireless Networks}
\label{sec:comparison}
Due to the many fundamental differences between classical cryptography
and information-theoretic security, it is useful to recognize what those
differences are and how they affect the choice of technology in a wireless
scenario. It is fair to state that classical cryptographic security
under the computational model offers the following advantages:
\begin{itemize}
\item there are so far no publicly-known, efficient attacks on public-key
systems such as RSA, and hence they are deemed secure for a large
number of applications;
\item very few assumptions are made about
the plaintext to be encoded, and security is provided on a
block-to-block basis, meaning as long as the cryptographic
primitive is secure, then every encoded block is secure;
\item Systems are widely deployed, technology is readily available and
inexpensive.
\end{itemize}
On the other hand, we must consider also the following disadvantages
of the computational model:
\begin{itemize}
\item Security is based on unproven assumptions regarding the
hardness of certain one-way functions. Plaintext is insecure if
assumptions are wrong or if efficient attacks are developed;
\item   In general there are no precise metrics or absolute comparisons
between various cryptographic primitives that show the trade off
between reliability and security as a function of the block length
of plaintext and ciphertext messages - in general, the security of
the cryptographic protocol is measured by whether it survives a
set of attacks or not;
\item   In general, these will not be information theoretically secure if the
communication channel between friendly parties and the eavesdropper
are noiseless, because the secrecy capacity of these application-layer
systems is zero;
\item State-of-the art key distribution schemes for wireless networks
based on the computational model require a trusted third party as well as
complex protocols and system architectures~\cite{aziz93privacy}.
\end{itemize}
The advantages of physical layer security under the information-theoretic (perfect)
security models can be summarized as follows:
\begin{itemize}
\item   No computational restrictions
are placed on the eavesdropper;
\item Very precise statements can be
made about the information that is leaked to the eavesdropper as a
function of channel quality and blocklength of the messages~\cite{BBCM95};
\item   Has been realized in practice through quantum key
distribution~\cite{bb84};
\item
   In theory, suitably long codes used
for privacy amplification can get exponentially close to perfect
secrecy~\cite{BBCM95};
\item Instead of distributing keys it is possible to generate on-the-fly as many
secret keys as desired.
\end{itemize}
In contrast, we have to take into consideration the following disadvantages
of information-theoretic security:
\begin{itemize}
\item   Information-theoretic
security is an average-information measure.  The system can be
designed and tuned for a specific level of security - e.g. with
very high probability a block will be secure, but it may not be
able to guarantee security with probability 1;
\item   Requires
assumptions about the communication channels that may not be
accurate in practice.  In many cases one would make very
conservative assumptions about the channels.  This will likely
result in low secrecy capacities and low secret-key or -message
exchange rates.  This gives extremely high security and
reliability, but at low communication rates;
\item   A few systems (e.g Quantum Key Distribution)
are deployed but the technology is not as widely
available and is expensive;
\item A short secret key is still required for authentication
~\cite{Maurer:93}.
\end{itemize}

In light of the brief comparisons above, it is likely that any
deployment of a physical-layer security protocol in a classical
system would be part of a "layered security" solution where
security is provided at a number of different layers, each with a
specific goal in mind.  This modular approach is how virtually all
systems are designed today, so in this context, physical-layer
security provides an additional layer of security that does not
exist today in classical systems.

\section{Conclusions}
\label{sec:conclusions}

We provided a preliminary characterization of the outage secrecy
capacity of wireless channels with quasi-static fading.
Specifically, we assumed that Alice --- having access to the CSI of
the main channel only --- chooses a target secrecy rate $R_s$
(without knowing the wiretap channel) and we investigated the outage
probability defined as $\prob(R_s>C_s)$. Our results reveal that (a)
perfectly secure communication over wireless channels is possible
even when the eavesdropper has a better average SNR than the
legitimate partners, and (b) the outage secrecy capacity of wireless
channels can actually be higher than the secrecy capacity of a
Gaussian wiretap channel with the same averaged SNRs $\gamma_M$ and
$\gamma_W$. Furthermore, we analyzed the impact of imperfect channel
state information on the outage probability and the outage secrecy
capacity. In particular, we have demonstrated that even in the
presence of imperfect CSI it is possible to convey information in an
almost secure manner, that is, with an average secure throughput
substantially greater than the average leaked throughput.

Suppose now that Alice has access to CSI on both the main channel
and the eavesdropper's channel. This is the case, for example in a
Time Division Multiple Access (TDMA) environment, when Eve is not
a covert eavesdropper, but simply another user interacting with
the wireless network, thus sending communication signals that
allow Alice to estimate the CSI of the channel between them. A
natural way for Alice to exploit the available CSI on both
channels to achieve secrecy is by transmitting useful symbols to
Bob only when the instantaneous SNR values are such that the
instantaneous secrecy capacity is strictly positive
($\gamma_M>\gamma_W$).

This observation thus suggests an {\it
opportunistic} secret key agreement scheme for wireless networks
--- even when the outage probability is very high, the available
secrecy capacity is still likely to enable Alice and Bob to
generate an (information-theoretically secured) encryption key
that could then be used to secure the data exchange while the
system is in outage of secrecy capacity. Implementing such a scheme
is the goal of Part II of this paper.

\bibliographystyle{IEEEtran}

\end{document}